\documentclass[acmsmall, screen, nonacm]{acmart}

\AtBeginDocument{%
  }

\usepackage{algorithm}      
\usepackage{algpseudocode}  
\usepackage{enumitem}
\usepackage{float}
\usepackage{placeins} 
\usepackage{booktabs}    
\usepackage{tabularx}
\usepackage{xcolor}
\usepackage{multirow}
\usepackage{array}
\usepackage[table]{xcolor}
\usepackage{colortbl}
\usepackage{float}
\usepackage{xspace}
\usepackage{graphicx}
\usepackage{xspace}

\newcommand{\ourmodel}{\textsc{SA-CAISR}\xspace}

\begin{document}
\let\WriteBookmarks\relax
\def\floatpagepagefraction{1}
\def\textpagefraction{.001}
\title{\ourmodel: Stage-Adaptive and Conflict-Aware Incremental Sequential Recommendation}

\author{Xiaomeng Song}
\affiliation{%
  \institution{School of Computer Science and Technology, Shandong University}
  \city{Qingdao}
  \country{China}
}
\email{xiaomengsong@sdu.edu.cn}

\author{Xinru Wang}
\affiliation{%
  \institution{School of Computer Science and Technology, Shandong University}
  \city{Qingdao}
  \country{China}}
\email{202535331@mail.sdu.edu.cn}

\author{Hanbing Wang}
\affiliation{%
  \institution{Michigan State University}
  \city{East Lansing}
  \state{MI}
  \country{USA}}
\email{hanbingwang01@gmail.com}

\author{Hongyu Lu}
\affiliation{%
  \institution{WeChat, Tencent}
  \city{Shenzhen}
  \country{China}}
\email{luhy94@gmail.com}

\author{Yu Chen}
\affiliation{%
  \institution{WeChat, Tencent}
  \city{Shenzhen}
  \country{China}}
\email{nealcui@tencent.com}

\author{Zhaochun Ren}
\affiliation{%
  \institution{Leiden University}
  \city{Leiden}
  \country{Netherlands}}
\email{z.ren@liacs.leidenuniv.nl}

\author{Zhumin Chen}
\authornote{Corresponding author.}
\affiliation{%
  \institution{School of Computer Science and Technology, Shandong University}
  \city{Qingdao}
  \country{China}}
\email{chenzhumin@sdu.edu.cn}

\renewcommand{\shortauthors}{Xiaomeng Song et al.}

\begin{abstract}
Sequential recommendation (SR) aims to predict a user's next action by learning from their historical interaction sequences. In real-world applications, these models require periodic updates to adapt to new interactions and evolving user preferences. While incremental learning methods facilitate these updates, they face significant challenges. Replay-based approaches incur high memory and computational costs, and regularization-based methods often struggle to discard outdated or conflicting knowledge.

To overcome these challenges, we propose \ourmodel, a \underline{S}tage-\underline{A}daptive and
\underline{C}onflict-\underline{A}ware 
\underline{I}ncremental 
\underline{S}equential 
\underline{R}ecommendation framework. As a buffer-free framework, \ourmodel operates using only the old model and new data, directly addressing the high costs of replay-based techniques. \ourmodel introduces a novel Fisher-weighted knowledge-screening mechanism that dynamically identifies outdated knowledge by estimating parameter-level conflicts between the old model and new data, selectively removing obsolete knowledge while preserving compatible historical patterns. This dynamic balance between stability and adaptability allows our method to achieve state-of-the-art performance in incremental SR. Specifically, \ourmodel improves Recall@20 by 2.0\% on average across datasets, while reducing memory usage by 97.5\% and training time by 46.9\% compared to the best baseline. This efficiency allows real-world systems to rapidly update user profiles with minimal computational overhead, ensuring more timely and accurate recommendations. Our implementation can be found at: \url{https://github.com/Togmm/SA-CAISR}
\end{abstract}




\maketitle

\section{Introduction}

Sequential Recommendation (SR) aims to capture the temporal evolution of user preferences by leveraging historical interaction sequences, offering more accurate predictions than traditional static methods~\citep{he2017neural, wang2019sequential}. 

In real-world applications, user behaviors are inherently non-stationary: new users and items continuously emerge, short-term trends coexist with long-term interests, and preferences drift over time~\citep{zhang2019survey,liu2024Integrating}. Recent studies emphasize that efficiently capturing these shifts from continuous data streams is critical for maintaining recommendation accuracy~\citep{Lee2025CapturingUI}.This dynamic nature often leads to significant distribution shifts between historical and future interactions, challenging the model's ability to generalize to out-of-distribution (OOD) samples~\citep{zhang2024Disentangled}. To address such shifts, SR models must be updated frequently. However, frequent retraining on the entire historical dataset is both time-consuming and computationally expensive, while naive incremental fine‑tuning on only the newly accumulated data typically leads to catastrophic forgetting of previously learned knowledge~\citep{cai2022reloop, lopezpaz2017gem, de2022survey}.

To address these challenges, Incremental Learning (IL) has been adopted to enable adaptive updates for SR models while preserving historical knowledge. Existing methods typically fall into two paradigms, yet both face fundamental limitations. 

\begin{itemize}
    \item Replay-based methods alleviate catastrophic forgetting by storing and replaying a subset of past interactions or their compressed representations~\citep{mi2020ader, mi2020man, zhang2024infer, buzzega2020dark, zhu2023reloop2}. While effective for stability, they incur substantial memory and computational overhead. Moreover, reliance on heuristics for sample selection often fails to retain interactions that are crucial for accurate long-term preference modeling, and the storage of raw user data raises privacy concerns, further restricting their deployments in sensitive domains~\citep{mi2020ader, zhang2024infer}. 
    
    \item Regularization-based approaches constrain parameter updates to mitigate forgetting~\citep{kirkpatrick2017overcoming, li2018learning, zhang2020graphsail}. They improve efficiency by avoiding explicit storage of historical data. Although efficient, these methods typically estimate importance exclusively from the old model's perspective~\citep{kirkpatrick2017overcoming, aljundi2018mas}. Consequently, they lack the mechanism to adaptively filter out outdated or conflicting knowledge, often over-constraining the model with obsolete information or failing to account for drastic distribution shifts~\citep{Liu2023linrec, Yoo2025pisa}. 
 
\end{itemize}

\begin{figure*}[!t]  
    \centering
    \includegraphics[width=0.8\textwidth]{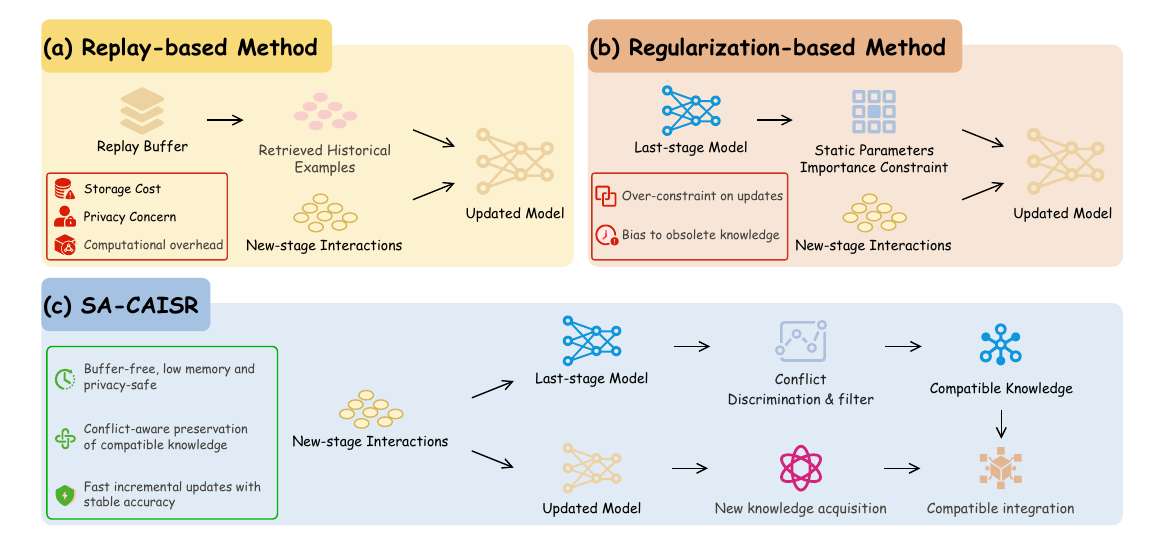}
    \caption{
    Overall illustration of three representative frameworks for incremental sequential recommendation.
    (a) Replay-based methods require storing and retrieving historical interaction sequences, leading to additional memory and retrieval overhead. 
    (b) Regularization-based methods constrain parameter updates based on fixed importance from the old model, lacking the ability to adaptively handle outdated or conflicting knowledge. 
    (c) The proposed \ourmodel framework introduces a Fisher-guided knowledge screening mechanism that selectively filters conflicting parameters, enabling fine-grained knowledge updating while preserving compatible historical information.
    }
    \Description{A diagram showing the comparison of methods.}
    \label{fig:method_comparison}
\end{figure*}

To overcome these limitations, our objective is to design a buffer-free framework that eliminates the memory and privacy burdens of replay methods, while equipping the model with the ability to actively identify and filter conflicting knowledge—a capability missing in traditional regularization approaches. To this end, we propose the 
\underline{S}tage-\underline{A}daptive and
\underline{C}onflict-\underline{A}ware 
\underline{I}ncremental 
\underline{S}equential 
\underline{R}ecommendation (\ourmodel) 
framework. As illustrated in Figure~\ref{fig:method_comparison}, unlike replay methods, \ourmodel is buffer-free, thereby preserving privacy and operational efficiency. Distinct from traditional regularization, it introduces a Fisher-weighted knowledge-screening mechanism. This mechanism computes Fisher information on new data to quantify parameter-level conflicts, selectively masking parameters in the old model. This enables the model to dynamically discard outdated or conflicting knowledge while strictly preserving compatible historical patterns. 
To ensure the updated model retains access to valid historical signals, we further integrate an InfoNCE-based consistency loss. Experimental results validate the superior efficacy of \ourmodel, demonstrating that it outperforms all baselines—including resource-intensive methods that utilize the complete historical data. Specifically, \ourmodel improves Recall@20 by 2.0\%, MRR@20 by 2.1\%, and NDCG@20 by 2.1\% on average across datasets. 
Beyond accuracy, the framework exhibits exceptional efficiency. While emerging lightweight architectures~\citep{zhang2024lightweight, wu2025lightweight} primarily focus on optimizing inference latency via structural designs, \ourmodel tackles the computational bottleneck of continuous training. Consequently, it reduces memory usage by 97.5\% and training latency by 46.9\% compared to representative baselines, striking an optimal balance between high-performance recommendation and computational scalability.

The main strengths of \ourmodel can be summarized as follows:

\begin{itemize}
    \item We propose a framework that operates using only the old model and new data, eliminating the need for interaction replay buffers and significantly reducing resource overhead.

    \item We introduce a novel mechanism that dynamically identifies and filters conflicting knowledge based on parameter-level importance, effectively balancing stability and plasticity.

    \item We introduce a dual-objective framework combining cross-entropy for new knowledge acquisition with an InfoNCE-based consistency constraint to preserve compatible historical patterns.

    \item Through comprehensive experimentation on four publicly available datasets, \ourmodel demonstrates significant performance and efficiency enhancements over state-of-the-art baseline models, making it highly suitable for large-scale, real-time recommendation scenarios. 

\end{itemize}

\section{Related Work}

Sequential recommendation (SR) aims to model users’ dynamic interests by leveraging their chronological interaction histories to predict the next item. Early SR research focused on traditional methods such as FPMC~\citep{rendle2010fpmc}. Later, neural models became dominant, including RNN-based approaches like GRU4Rec~\citep{hidasi2016gru4rec}, Transformer-based methods such as SASRec~\citep{kang2018sasrec} and BERT4Rec~\citep{sun2019bert4rec}. 
To handle complex sequences and leverage external knowledge, efficient State Space Models (SSMs)~\citep{yang2024uncovering} and LLM-based recommendation~\citep{wang2024rethinking, rajput2023recommender} have emerged as rapidly growing directions. Notably, recent studies have extended these paradigms to generative retrieval and data-efficient tuning~\citep{lin2024data, li2023generative}, significantly enhancing representation capabilities but also increasing the computational cost of model updates.
As user preferences and item pools evolve continuously in real-world applications, SR models must be updated frequently. This necessity gives rise to Incremental Sequential Recommendation (ISR), which focuses on enabling continual adaptation while mitigating catastrophic forgetting. Existing methods can be broadly categorized into two mainstream families: Replay-based approaches and Regularization-based approaches.

\subsection{Replay-based Methods}
Replay-based methods, also known as memory-based approaches, alleviate catastrophic forgetting by explicitly storing and re-training on a subset of historical data alongside new interactions. The core challenge in this domain lies in selecting the most informative samples to maximize knowledge retention under strict memory budgets.
Early representative approaches, such as ADER~\citep{mi2020ader} and MAN~\citep{mi2020man}, focus on interaction-level sampling. ADER, for instance, employs a herding-based strategy to maintain a diverse reservoir of user-item interactions that best represent the historical distribution. 

More recent efforts have evolved from simple sampling to structure-aware and feedback-loop mechanisms. In the context of graph-based recommendation, Inv-Degree~\citep{ahrabian2021structure} argues that random sampling disrupts graph connectivity. Instead, it introduces a topology-preserving selection mechanism that prioritizes high-degree nodes and critical structural paths to maintain the integrity of the interaction graph during updates. 
Furthermore, methods like ReLoop~\citep{cai2022reloop} and ReLoop2~\citep{zhu2023reloop2} extend the replay paradigm by integrating a self-correcting feedback loop. These frameworks maintain a separate memory buffer for error compensation, enabling the system to identify during online inference the samples on which the model performed poorly or was highly uncertain, and to “replay” them in subsequent training.

Despite their effectiveness, replay-based approaches face inherent scalability and ethical hurdles. The storage of historical data leads to a linear growth in memory costs and computational overhead for retrieval. More critically, retaining user interaction logs—even in sampled forms—raises significant privacy concerns and complicates compliance with data protection regulations~\citep{Arslan2019Agem, xuPrivacy2016}. In contrast, our proposed \ourmodel eliminates the need for a replay buffer entirely, relying instead on efficient model-centric constraints.

\subsection{Regularization and Distillation Methods}
To avoid the storage costs of replay, this category of methods focuses on constraining the optimization process to preserve knowledge. These approaches can be divided into parameter-level constraints, output-level distillation, and representation-level contrastive learning.

Classic parameter-level regularization methods, such as Elastic Weight Consolidation (EWC)~\citep{kirkpatrick2017overcoming}, aim to protect critical model parameters. EWC identifies these weights by evaluating the sensitivity of the model's performance on data from previous stages to parameter updates, effectively penalizing changes to important weights.
In the domain of recommendation, these ideas have been extended to preserve structural knowledge. GraphSAIL~\citep{zhang2020graphsail} enforces both local and global structural constraints on the embedding space to maintain historical user-item relationships. Similarly, SAIL-PIW~\citep{wang2023sailpiw} introduces personalized imitation weights, adaptively adjusting the regularization strength for each user based on their interaction frequency and preference volatility.

Beyond direct parameter constraints, Knowledge Distillation (KD) and Contrastive Learning (CL) are widely used to maintain functional consistency. Learning without Forgetting (LwF)~\citep{li2018learning} is a seminal work using KD to align the output probabilities of the new model with the old model. In ISR, methods like IMSR~\citep{wang2023imsr} utilize KD to preserve existing multi-interest representations while allowing the model to learn new interest vectors.
Concurrently, Contrastive Learning has been adopted to enhance representation robustness. While early frameworks like CL4SRec~\citep{xie2022cl4srec} and CoSeRec~\citep{LiuCoSeRec2021} rely on basic data augmentations, recent studies have advanced towards capturing complex structural patterns. Notably, \citep{lee2025contrastive} introduces a hierarchical contrastive framework to align user preferences across different granularities. 
In the incremental setting, this paradigm is further adapted for stability. Approaches such as LWC-KD~\citep{wang2021LWC-KD} leverage contrastive objectives to enforce semantic consistency between old and new models, ensuring that updated representations do not deviate significantly from historical knowledge.

While these methods avoid storing past data, a common limitation remains: they often treat the historical model (the “teacher”) as a fixed reference, constraining the new model to match it without explicitly modeling distribution shift or conflict between old and new knowledge~\citep{Sebastian2019UnifyingBayesian}.
Recent advances in sequential recommendation have shifted towards disentangling complex user intents to isolate valid patterns from noise or conflicting signals~\citep{adsr2025}. While such methods leverage Large Language Models (LLMs) to achieve high-quality disentanglement through semantic reasoning, they often incur substantial computational overhead unsuitable for real-time incremental updates.
It is worth noting that a recent concurrent work, FGGM~\cite{tan2026fggm}, also explores Fisher-guided gradient masking to address similar challenges in general continual learning. 
However, while sharing high-level similarities in leveraging Fisher Information, FGGM focuses on static classification tasks. 
In contrast, our \ourmodel is specifically tailored for sequential recommendation, addressing the unique challenge of user preference drift through a stage-adaptive and conflict-aware mechanism.
Our method, \ourmodel, addresses this by integrating the conflict-awareness of parameter regularization with the semantic consistency of contrastive learning. Rather than distilling from a static teacher, we employ a Fisher-guided masking mechanism to dynamically filter out conflicting parameters from the reference model. This creates a "purified" teacher for the InfoNCE loss, ensuring that the model preserves only compatible historical knowledge while freely adapting to new trends.

\subsection{Summary}
 In summary, existing ISR methods struggle to balance adaptability and efficiency. Regularization-based approaches are lightweight but typically lack explicit, conflict-aware forgetting mechanisms, which can over-constrain updates under distribution shift, whereas replay-based methods offer stability at the cost of scalability, memory, and privacy. Our \ourmodel addresses these challenges via two components: (1) a Fisher-weighted knowledge-screening mechanism that selectively filters or suppresses parameters conflicting with new data while preserving compatible patterns, and (2) an InfoNCE-based contrastive consistency constraint that stabilizes representation alignment during adaptation. Together, these components enable effective adaptation to evolving user behaviors while mitigating the influence of obsolete knowledge, achieving improved performance with minimal overhead.

\section{Preliminary}
This section establishes the mathematical notations and conceptual frameworks necessary for understanding the proposed approach. We structure the preliminaries into three key components: the formal definition of the Sequential Recommendation (SR) task, the architecture of Transformer-based models in this context, and the formulation of the Incremental Learning setting.

\subsection{Problem Statement and Formulation}
We first formalize the Sequential Recommendation problem. 
Let $\mathcal{U} = \{u_1, u_2, ...,u_{|\mathcal{U}|}\}$ denote a set of users and $\mathcal{I} = \{i_1, i_2, ...,i_{|\mathcal{I}|}\} $ denote a set of items. For each user $u \in \mathcal{U}$, the historical interaction sequence is defined as:
\begin{equation}
S_u = [i_1, i_2, \dots, i_t], \quad i_k \in \mathcal{I}, \, 1 \le k \le t,
\end{equation}
where $i_k$ denotes the item interacted at time step $k$. Given the interaction sequence $S_u$ of user $u$, sequential recommendation models aim to predict the next item $i_{t+1}$ the user is most likely to interact with from $\mathcal{I}$. 

The training objective is to maximize the log-likelihood of the true next item:
\begin{equation}
\mathcal{L}_{\mathrm{SR}} = - \sum_{u \in \mathcal{U}} \log P(i_{t+1} \mid S_u).
\end{equation}

\subsection{Transformer-based Sequential Recommendation}

We next examine the application of Transformer architectures to the SR problem. Unlike Recurrent Neural Networks (RNNs) ~\citep{hidasi2016gru4rec, Hidasi2018rnn4sr} or Convolutional Neural Networks (CNNs)~\citep{tang2018caser, yuan2019nir, YAKHCHI2022CAN}, Transformer-based models (such as SASRec~\citep{kang2018sasrec}) leverage the self-attention mechanism to capture complex dependencies within user interaction sequences. 

In this formulation, the user sequence $S_u = [i_1, i_2, \ldots, i_t]$ is converted into a sequence of embedding vectors:
\begin{equation}
\mathbf{E} = [\mathbf{e}_{i_1}, \mathbf{e}_{i_2}, \ldots, \mathbf{e}_{i_t}], 
\quad \mathbf{e}_{i_k} \in \mathbb{R}^d,
\end{equation}
where $\mathbf{e}_{i_k}$ denotes the embedding of item $i_k$ and $d$ is the embedding dimension. To preserve the order information within the sequence, positional embeddings are added:
\begin{equation}
\mathbf{H}^0 = \mathbf{E} + \mathbf{P},
\end{equation}
where $\mathbf{H}^0$ represents the input matrix and $\mathbf{P}$ is the positional encoding matrix.

At the $\ell$-th Transformer layer, the hidden representation is updated through the multi-head self-attention (MHSA) mechanism followed by a feed-forward network (FFN):
\begin{equation}
\mathbf{H}^\ell = \mathrm{FFN}(\mathrm{MHSA}(\mathbf{H}^{\ell-1})),
\end{equation}
where MHSA computes attention weights as:
\begin{equation}
\mathrm{Attention}(\mathbf{Q}, \mathbf{K}, \mathbf{V}) 
= \mathrm{softmax}\!\left(\frac{\mathbf{Q}\mathbf{K}^\top}{\sqrt{d}}\right)\mathbf{V}.
\end{equation}

The final hidden representation at the last position, 
$\mathbf{h}_t = \mathbf{H}^L_t$, is used to predict the next item:
\begin{equation}
\hat{i}_{t+1} = \arg\max_{i \in I} 
\; \mathrm{softmax}(\mathbf{h}_t^\top \mathbf{e}_i).
\end{equation}

The model is trained by maximizing the log-likelihood of the true next item:
\begin{equation}
L_{\mathrm{Trans}} = - \sum_{u \in U} 
\log P(i_{t+1} \mid S_u),
\end{equation}
where $P(i_{t+1} \mid S_u) = \mathrm{softmax}(\mathbf{h}_t^\top \mathbf{e}_{i_{t+1}})$.

\subsection{Incremental Learning for Sequential Recommendation}
Traditional SR models are typically trained on static datasets, where all user–item interaction sequences are available prior to training. However, real-world online systems operate in non-stationary environments where user interests and item distributions evolve continuously. The constant influx of new users, items, and interactions renders the cost of retraining models from scratch prohibitively expensive. This dynamic nature motivates the Incremental Sequential Recommendation (ISR) problem, a paradigm that extends the standard SR setting into a continual learning framework.

Formally, we segment the interaction stream into temporal stages $\{\mathcal{D}_1,\mathcal{D}_2,\dots,\mathcal{D}_M\}$, where each stage aggregates interactions within a fixed time window. For stage $m$, we write
$\mathcal{D}_m = \{(S_u^m[1\!:\!t],\, i_{t+1}^m)\}$,
where $u \in \mathcal{U}$, $i_{t+1}^m \in \mathcal{I}$, and $S_u^m[1\!:\!t] = [i_1^m, i_2^m, \dots, i_t^m]$ denotes the prefix of user $u$’s interaction sequence within stage $m$ up to position $t$. Given the context $S_u^m[1\!:\!t]$, the model predicts the next item $i_{t+1}^m$ over the (possibly growing) item universe $\mathcal{I}$. Unlike conventional SR that learns a static mapping $f_\theta(S_u) \to i_{t+1}$ on a fixed dataset, ISR continuously refines the model as new data $\mathcal{D}_m$ arrive, adapting to new patterns while retaining useful knowledge from previous stages to mitigate catastrophic forgetting. 

Let $\theta^m$ denote the model parameters after training on stage $m$. 
Upon progressing to the subsequent stage, the model is initialized with $\theta^{m-1}$ and adapted using the incoming data $\mathcal{D}^m$. 
The general optimization objective can be formulated as:
\begin{equation}
L_{\mathrm{ISR}}^m = L_{\mathrm{new}}^m + \mathcal{R}(\theta^m, \theta^{<m}),
\end{equation}
where $L_{\mathrm{new}}^m$ is the standard recommendation loss (e.g., cross-entropy) on $\mathcal{D}^m$, 
and $\mathcal{R}(\theta^m, \theta^{<m})$ serves as a regularization term to preserve previously acquired knowledge.

Different ISR approaches are distinguished by their instantiation of $\mathcal{R}(\cdot)$: regularization-based methods (e.g., EWC) constrain updates on parameters identified as critical for prior tasks; replay-based approaches maintain a memory buffer of representative samples for joint training. Consequently, ISR can be viewed as solving the following optimization problem:
\begin{equation}
\theta^m = \arg\min_\theta \big[L_{\mathrm{new}}^m + \mathcal{R}(\theta, \theta^{<m})\big],
\end{equation}
achieving a balance between plasticity for new user preferences and stability for preserving historical knowledge.

\section{Method}

\begin{figure}[t]
    \centering
    \includegraphics[width=1\textwidth]{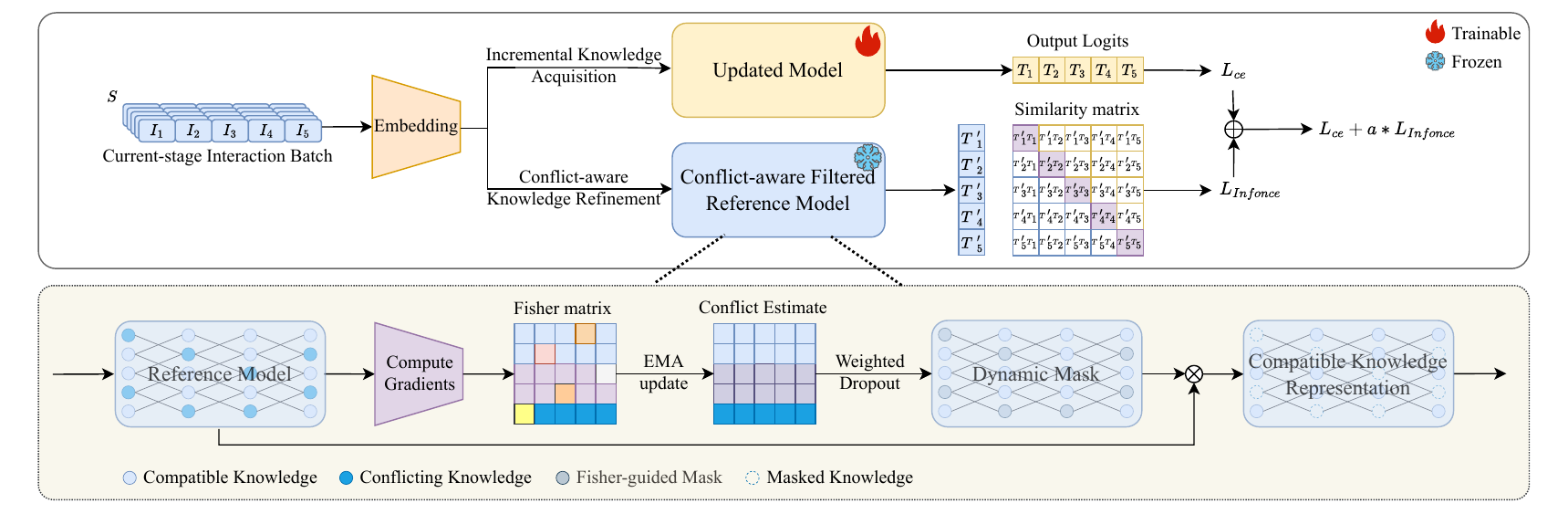}
    \caption{
        Overall architecture of \ourmodel. 
        The Current-stage Interactions are processed through two parallel branches: 
        (1) the \textit{training branch}, which learns new knowledge through cross-entropy loss and 
        (2) the \textit{reference branch}, which estimates Fisher information from the reference model to compute parameter conflicts and obtain a Compatible Knowledge representation via selective masking.
        The two branches are aligned using an InfoNCE-based consistency loss, forming a joint objective that balances new knowledge acquisition and preservation of compatible historical knowledge. 
        Fisher-guided selective masking and contrastive regularization together enable stable and efficient incremental adaptation.
    }
    \Description{A diagram showing the framework.}
    \label{fig:framework}
\end{figure}

In this section, we present \ourmodel. As illustrated in Figure~\ref{fig:framework}, our framework addresses the incremental learning challenge by augmenting a standard sequential encoder with two novel components: a \textbf{Conflict-Aware Knowledge Screening} mechanism and a \textbf{Consistency Preservation} module. We build on a pure attention-based sequential backbone (SASRec) and do not use LLMs. Our online/near-real-time setting prioritizes end-to-end latency and throughput; LLMs’ token-level generation and long-context computation raise inference/training latency, challenging second-level SLAs and high-QPS serving. To isolate the effect of our incremental strategy, we keep the SASRec-consistent backbone and modality (pure sequential signals, no external semantics), avoiding capacity/modality confounders and aligning with low-latency, production needs. This attention-only setup is also common in concurrent sequential/incremental recommendation work under similar deployment constraints, supporting fair comparison and practical applicability. We use the following terminology for each training stage $t$. The parameters $W_{t-1}$ represent the model state obtained after completing stage $t-1$. The reference model $f_{\mathrm{ref}}(\cdot; W_{\mathrm{ref}})$ is a frozen copy of the previous-stage model, initialized as $W_{\mathrm{ref}} \leftarrow W_{t-1}$; it carries historical knowledge and is not updated by gradient descent within stage $t$. The Fisher-filtered reference model $f_{\mathrm{ref}}(\cdot; \widetilde{W}_{\mathrm{ref}})$ is obtained by applying a Fisher-guided probabilistic parameter mask $m$ (resampled per batch) to the reference model to suppress parameters conflicting with the new data distribution, yielding $\widetilde{W}_{\mathrm{ref}} = W_{\mathrm{ref}} \odot m$. The updated model $f_{\mathrm{updated}}(\cdot; W_{\mathrm{updated}})$ is the trainable model for stage $t$, initialized from the previous stage as $W_{\mathrm{updated}} \leftarrow W_{t-1}$, and optimized on the current stage’s data to acquire new knowledge.

\subsection{Conflict-Aware Knowledge Screening}
\label{sec:fisher} 

To prevent catastrophic forgetting in incremental learning, it is insufficient to simply retain historical weights; we must identify which parameters encode valid historical knowledge and which have become outdated due to shifting user preferences. To this end, we propose a \textit{Fisher-guided screening mechanism}. This approach dynamically filters out conflicting or outdated information from the reference model, ensuring that the updated model is only constrained to align with valid historical patterns.

\subsubsection{Fisher Information Estimation} 

We utilize the Fisher Information Matrix (FIM) to quantify the sensitivity of the reference model's parameters to the incoming data distribution. The rationale for employing FIM in this scenario lies in its ability to serve as a rigorous proxy for knowledge conflict. In statistical learning, Fisher Information measures how strongly a parameter influences the likelihood of the observed data. Intuitively, if a parameter exhibits high Fisher information on the new data, it implies that the historical knowledge encoded in that parameter is highly unstable under the current distribution. Therefore, we use FIM to identify parameters where the gradient of the log-likelihood fluctuates significantly, signaling a discrepancy between the reference model’s memory and the emerging user preferences.

Formally, let the data distribution be \(p(x \mid \theta)\), where \(\theta\) denotes model parameters and \(x\) is an observed sample. The FIM is defined as:

\begin{equation}
F(\theta) = \mathbb{E}_{x \sim p(x \mid \theta)} \Bigg[ \Bigg( \frac{\partial}{\partial \theta} \log p(x \mid \theta) \Bigg) \Bigg( \frac{\partial}{\partial \theta} \log p(x \mid \theta) \Bigg)^\top \Bigg],
\end{equation}

Here, \(\log p(x \mid \theta)\) represents the log-likelihood function. The Fisher Information Matrix is positive semi-definite, and the diagonal entries capture the importance of individual parameters. However, computing the full matrix is computationally prohibitive for deep neural networks. Therefore, we adopt the diagonal approximation~\citep{james2020kfac}:  

\begin{equation}
F_i \approx \mathbb{E}_{x \sim \mathcal{D}} \Bigg[ \Bigg( \frac{\partial}{\partial \theta_i} \log p(x \mid \theta) \Bigg)^2 \Bigg],
\label{eq:fisher}
\end{equation}

where \(F_i\) denotes the Fisher information value for parameter \(\theta_i\), and \(\mathcal{D}\) is the training dataset. A larger \(F_i\) indicates that the parameter plays a more critical role in modeling the data distribution, and drastic updates to such a parameter are likely to cause significant interference with previously learned knowledge. 

In \ourmodel, Fisher information is reinterpreted as a measure of parameter conflict rather than static importance. This interpretation is justified by the fact that we compute the Fisher metric on the new data distribution using the frozen historical parameters. In this context, a large Fisher value implies that the parameter generates high-magnitude gradients, signaling that its current (historical) state is significantly suboptimal for the incoming data patterns. Consequently, this sensitivity serves as a direct proxy for the discrepancy—or conflict—between the model's historical memory and the emerging user preferences, thereby serving as the basis for knowledge filtering in the reference model.

As new data arrives, we compute gradients through this frozen reference model. A large \(F_i\) calculated on new data indicates that the parameter is highly sensitive to the current distribution and requires significant updating. This high sensitivity implies a conflict between the parameter's historical value and the new data patterns. Consequently, such parameters represent potentially outdated knowledge that should be screened out (See Algorithm~\ref{alg:incremental}, steps 5–8).

To stabilize these estimates and mitigate batch-wise noise, we employ an exponential moving average (EMA) update across batches~\citep{p2017adam}, which can be defined as:
\begin{equation}
F_i = \beta \cdot F_{i-1} + (1 - \beta) \cdot g_i^2,
\label{eq:fisher_updata}
\end{equation}
where $g_i$ denotes the gradient of parameter $\theta_i$ in the current batch and $\beta$ is the smoothing coefficient (See Algorithm~\ref{alg:incremental}, step 9).  

\subsubsection{Adaptive Masking Mechanism}

We leverage the derived conflict scores to impose a probabilistic mask on the reference model. The objective is to ``silence'' parameters that conflict with the current stage, thereby preventing the model from retaining incompatible historical patterns. Acknowledging that Fisher information magnitudes can vary significantly across different layers, we employ a layer-wise normalization strategy. Specifically, for each parameter $i$ within a specific layer (or parameter tensor) $l$, the Fisher value $F_i$ is normalized and scaled to determine the masking probability $p_i$:

\begin{equation}
p_{i} = \frac{F_{i} - \min(\mathcal{F}_l)}{\max(\mathcal{F}_l) - \min(\mathcal{F}_l) + \epsilon} \cdot p_{\max},
\label{eq:fisher_scaled}
\end{equation}

where $\mathcal{F}_l$ denotes the set of Fisher values associated with layer $l$, and $\epsilon$ is a small constant added for numerical stability. This normalization maps the conflict scores to a layer-adaptive scale within $[0, p_{\max}]$. Here, $p_{\max}$ is a hyperparameter that controls the maximum dropout probability (see Algorithm~\ref{alg:incremental}, step 11).
Following the principle of Dropout, masks are independently sampled for each parameter:
\begin{equation}
W^{ref}(i) =
\begin{cases}
0, &  p_i, \\
W^{ref}(i) / (1-p_i), &  1-p_i,
\end{cases}
\label{eq:fisher_dropout}
\end{equation}
where $W^{ref}(i)$ denotes the $i$-th parameter of the reference model and $p_i$ denotes the masking probability for the $i$-th parameter. This operation results in a filtered reference model that retains only consistent historical knowledge while suppressing conflicting parameters (Algorithm~\ref{alg:incremental}, step 12). 

It is important to note that this parameter modification is temporary. The scaling applied in Eq.~(\ref{eq:fisher_dropout}) is reverted, and the reference model is restored to its original frozen state immediately after the loss computation for the current batch (Algorithm~\ref{alg:incremental}, step 15). This ensures that the reference model remains a stable baseline throughout the epoch. Masks are resampled every batch to ensure robust coverage. Finally, to allow for long-term adaptation, the accumulated Fisher weights are decayed at the end of each epoch:

\begin{equation}
F_{i+1} \leftarrow \gamma \cdot F_i,
\label{eq:fisher_decay}
\end{equation}
where $\gamma$ is a decay factor that gradually weakens the accumulated importance from previous stages (See Algorithm~\ref{alg:incremental}, step 21).

\subsection{Consistency Preservation via Contrastive Alignment}
\label{sec:infonce}


Having screened out conflicting parameters, the next objective is to align the updated model with the valid historical knowledge preserved in the reference model. Instead of imposing rigid parameter-level constraints, we seek a flexible alignment in the representation space. Inspired by recent advances demonstrating that contrastive signals significantly enhance representation robustness against sequence uncertainty and noise~\citep{wang23srmc, zhang2024soft}, we employ the InfoNCE objective~\citep{xie2022cl4srec}. The core logic is to maximize the similarity between the updated representation and the consistent historical representation, while differentiating them from other items.

Formally, consider a batch of $N$ samples $\{x_i\}_{i=1}^N$. Let $\mathbf{h}_i$ denote the representation of sample $x_i$ produced by an anchor encoder, and let \(\mathbf{h}_i^+\) denote the corresponding positive representation produced by a positive encoder (serving as a teacher) for the same input. For efficiency, we adopt in‑batch negative sampling: the remaining samples in the batch serve as candidate negatives. The InfoNCE loss is:

\begin{equation}
\mathcal{L}_{\text{InfoNCE}} = - \sum_{i=1}^{N} \log \frac{\exp(\text{sim}(\mathbf{h}_i, \mathbf{h}_i^+)/\tau)}{\sum_{j=1}^{N} \exp(\text{sim}(\mathbf{h}_i, \mathbf{h}_j)/\tau)},
\end{equation}

where $\mathrm{sim}(\cdot,\cdot)$ is a similarity function (e.g., cosine or inner product) and $\tau$ is a temperature. This objective maximizes the agreement between each anchor and its positive while separating it from negatives. A common variant replaces the full‑batch denominator with only the hardest top‑$k$ negatives to emphasize the most confusable samples.

In \ourmodel, the anchor representation is $\mathbf{h}_i = f_{\text{upd}}(x_i)$ from the updated model. The positive is produced by a dynamically filtered reference (teacher) model: $\mathbf{h}_i^{+} = f_{\text{ref}}(x_i; \widetilde{W}_{\text{ref}})$, where $\widetilde{W}_{\text{ref}} = W_{\text{ref}} \odot m$ and $m$ is a Fisher‑guided, probabilistic mask resampled per batch on new‑stage data to suppress conflicting parameters and preserve compatible historical knowledge. For negatives, we use in‑batch candidates and select the top‑$k$ hardest for each anchor based on current similarity, yielding
\begin{equation}
\mathcal{L}_{\text{InfoNCE}}^{(k)}
=
-\frac{1}{N}
\sum_{i=1}^{N}
\log
\frac{\exp\big(\mathrm{sim}(\mathbf{h}_i, \mathbf{h}_i^{+})/\tau\big)}
{\exp\big(\mathrm{sim}(\mathbf{h}_i, \mathbf{h}_i^{+})/\tau\big)
+
\sum_{j \in \mathcal{N}^{(k)}_i}
\exp\big(\mathrm{sim}(\mathbf{h}_i, \mathbf{h}_j)/\tau\big)}\,,
\label{eq:infonce}
\end{equation}
where $\mathcal{N}^{(k)}_i$ indexes the top‑$k$ hard negatives for anchor $i$. When aligning score vectors instead of pure embeddings, both branches are first mapped through a shared projection into a common representation space before computing similarity. The top‑$k$ strategy improves fine‑grained discrimination and reduces computation versus using all batch negatives; however, too small a $k$ may omit informative negatives, so $k$ is tuned to balance discriminability and stability (see Algorithm\ref{alg:incremental}, step 14, and ablations).

\subsection{Model Optimization}

The ultimate objective of incremental sequential recommendation is to acquire new user interaction patterns while preserving essential historical knowledge. To achieve this balance, we integrate the two complementary objectives: Cross-Entropy (CE) for learning new knowledge and InfoNCE for preserving old knowledge.  

\noindent \textbf{New Knowledge Acquisition}. 
To capture emerging user interests, we train the updated model on the current stage dataset $\mathcal{D}^t$ using the Cross-Entropy (CE) loss:

\begin{equation}
\mathcal{L}_{\text{CE}}^t = - \sum_{(S_u^t, i_{t+1}) \in \mathcal{D}^t} \log P(i_{t+1} \mid S_u^t), \label{eq:ce}
\end{equation}

where $(S_u^t, i_{t+1})$ denotes the historical sequence and the next-item target for user $u$ at the current stage, and $P(i_{t+1} \mid S_u^t)$ is the probability predicted by the sequential recommendation model. Minimizing $\mathcal{L}_{\text{CE}}^t$ enables the model to learn new interaction patterns efficiently. 

\noindent \textbf{Joint Objective}. 
Relying solely on CE risks catastrophic forgetting. Therefore, we integrate the conflict-aware InfoNCE loss to explicitly constrain the optimization trajectory. The total loss is defined as:

\begin{equation}
\mathcal{L}^t = \mathcal{L}_{\text{CE}}^t + \alpha \cdot \mathcal{L}_{\text{InfoNCE}}^t,
\label{eq:loss_tol}
\end{equation}

where $\alpha$ is a hyperparameter that controls the relative importance of historical knowledge retention. A larger $\alpha$ emphasizes alignment with the reference model and thus strengthens memory preservation, while a smaller $\alpha$ prioritizes adapting to newly emerging behaviors (See Algorithm~\ref{alg:incremental}, step 17).  

This joint objective allows the model to update itself in a balanced manner:  
\begin{itemize}
    \item $\mathcal{L}_{\text{CE}}^t$ ensures that the model captures the most recent user preferences based on the current stage dataset $\mathcal{D}^t$.  
    \item $\mathcal{L}_{\text{InfoNCE}}^t$ constrains the model representations to remain consistent with the conflict-filtered outputs of the reference model, thereby preserving previously acquired knowledge and mitigating catastrophic forgetting.
\end{itemize}

In practice, the value of $\alpha$ can be tuned according to the data dynamics. For domains where user interests shift rapidly (e.g., news or short videos), a smaller $\alpha$ is preferable to allow rapid adaptation. Conversely, in domains with more stable preferences (e.g., e-commerce), a larger $\alpha$ is beneficial to ensure knowledge retention.  

By optimizing this weighted combination of CE and InfoNCE, \ourmodel effectively mitigates the stability–plasticity dilemma: it enables sufficient plasticity to learn novel information while maintaining stability against catastrophic forgetting, resulting in robust incremental sequential recommendation performance. 

\begin{algorithm}[t]
\caption{Training procedure of \ourmodel}
\label{alg:incremental}
\begin{algorithmic}[1]
\Statex \textbf{Input:} Previous-stage model $W_{t-1}$, new-stage dataset $\mathcal{D}_t$, Fisher weights $\mathcal{F}$, balancing factor $\alpha$, maximum masking probability $p_{\max}$, EMA factor $\beta$, decay factor $\gamma$
\Statex \textbf{Output:} Updated model $W_t$
\State $W^{updated} \gets W_{t-1}$, $W^{ref} \gets W_{t-1}$ (frozen), $\mathcal{F} \gets 0$
\For{each epoch}
    \For{each batch $B \in \mathcal{D}_t$}
        \State \textbf{(a) CE Loss Calculation:}
        \State Compute $\mathcal{L}_{CE}$ using Eq.~\ref{eq:ce}
        \State Compute gradients $\nabla_{W^{ref}}\mathcal{L}_{CE}$ on $B$ for Fisher calculation
        \State \textbf{(b) Fisher update:}
        \State Approximate Fisher information using Eq.~\ref{eq:fisher}
        \State Update parameter weights via Eq.~\ref{eq:fisher_updata}
        \State \textbf{(c) Knowledge filtering:}
        \State Compute parameter-wise masking probabilities from Fisher weights according to Eq.~\ref{eq:fisher_scaled}
        \State Apply dropout-style masking to $W^{ref}$ following Eq.~\ref{eq:fisher_dropout}
        \State \textbf{(d) Consistency regularization:}
        \State Compute $\mathcal{L}_{InfoNCE}$ between $W^{updated}$ and $W^{ref}_f$ using Eq.~\ref{eq:infonce}
        \State Restore $W^{ref}$ afterwards
        \State \textbf{(e) Joint Optimization:}
        \State $\mathcal{L}_{total} \gets \mathcal{L}_{CE} + \alpha \cdot \mathcal{L}_{InfoNCE}$
        \State Update $W^{updated}$ by minimizing $\mathcal{L}_{total}$
    \EndFor
    \State \textbf{(f) Fisher decay:}
    \State Apply decay to Fisher weights as specified in Eq.~\ref{eq:fisher_decay}
\EndFor
\State \textbf{Result:} $W_t = W^{updated}$
\end{algorithmic}
\end{algorithm}

\section{Experiments}
\label{sec:exp}

We aim to answer the following research questions (RQs):

\begin{itemize}[leftmargin=4em, labelsep=0.5em, align=left]
    \item[\textbf{(RQ1)}] How does \ourmodel perform in terms of recommendation accuracy and computational/memory efficiency compared with representative incremental sequential recommendation methods?
    \item[\textbf{(RQ2)}] How does \ourmodel handle different stage intervals and user preference dynamics in comparison with other methods?
    \item[\textbf{(RQ3)}] How do Fisher-guided conflict-aware screening and InfoNCE-based consistency collaboratively contribute to balancing stability and plasticity in incremental sequential recommendation?
\end{itemize}

\subsection{Experimental Settings}
\label{sec:expset}

\subsubsection{Datasets}
\label{sec:datasets}

To comprehensively evaluate the effectiveness of the proposed method, we conduct experiments on four widely used benchmark datasets with diverse temporal coverage and sequence characteristics: \textbf{DIGINETICA\footnote{\href{https://competitions.codalab.org/competitions/11161}{DIGINETICA: https://competitions.codalab.org/competitions/11161}}}~\citep{diginetica2016}, \textbf{YOOCHOOSE\footnote{\href{https://www.kaggle.com/datasets/chadgostopp/recsys-challenge-2015}{YOOCHOOSE: https://www.kaggle.com/datasets/chadgostopp/recsys-challenge-2015}}}~\citep{yoochoose2015}, \textbf{Gowalla\footnote{\href{https://snap.stanford.edu/data/loc-Gowalla.html}{Gowalla: https://snap.stanford.edu/data/loc-Gowalla.html}}}~\citep{gowalla2011}, and \textbf{Amazon Sports\footnote{\href{https://nijianmo.github.io/amazon/index.html}{Amazon Sports: https://nijianmo.github.io/amazon/index.html}}}~\citep{amazon2019}. 
Table~\ref{tab:dataset_stats} summarizes the detailed statistics of all datasets, including the number of users, items, and interactions, as well as their temporal spans and sequence length distributions. These dataset are selected to cover a wide spectrum of user behaviors and temporal dynamics, providing a solid foundation for evaluating the robustness and adaptability of our model under different continual learning settings.

\begin{itemize}
    \item DIGINETICA and YOOCHOOSE are session-based e-commerce datasets collected from real-world online shopping platforms, where user sessions are typically short and exhibit strong temporal drift due to rapidly changing interests. 
    \item Gowalla is a location-based social network dataset containing timestamped user check-in sequences collected from February 2009 to October 2010.The dataset reflects long-term user mobility behaviors with relatively stable but gradually evolving preferences over months.
    \item Amazon Sports is a subset of the Amazon review corpus in the Sports and Outdoors domain, covering user-item interactions from approximately 1996 to 2018.It represents a long-term recommendation scenario with multi-year interaction histories and slow preference shifts.
\end{itemize}

Together, these datasets span from short-term to long-term temporal ranges and from sparse to dense sequential interactions, enabling a comprehensive evaluation of how well our method balances plasticity and stability across different temporal granularities. 

For DIGINETICA and YOOCHOOSE, we follow the preprocessing strategy of ADER~\citep{mi2020ader}. To simulate a continual learning scenario, the DIGINETICA dataset is split into multiple stages by week, while YOOCHOOSE is divided by day. Among these, 16 stages are selected for each dataset to form the experimental setup. For Gowalla and Amazon Sports, we adopt the splitting strategy used in~\citep{zhang2020graphsail, Lee2025CapturingUI}. The entire dataset is partitioned into 5 blocks. The first block (\(D_0\)) consists of the earliest 60\% of the data and serves as the base block for pretraining before continual learning begins. The remaining 40\% of the data is equally divided into four incremental blocks (\(D_1\) to \(D_4\)) according to temporal order. Within each block of every dataset, we apply a \textit{k-core} filtering strategy to remove sessions shorter than a given threshold \(k\). Specifically, we set \(k=2\) for DIGINETICA and YOOCHOOSE, \(k=10\) for Gowalla, and \(k=3\) for Amazon Sports, where the choice of \(k\) depends on the average session length of each dataset.

Adhering to the training protocol of ADER~\citep{mi2020ader}, the model at each stage is initialized using parameters from the preceding stage. The data for the current stage is partitioned into 90\% for training and 10\% for validation, while the complete dataset of the subsequent stage is utilized for testing. To ensure evaluation fairness and consistency, any items in the test set that were not observed in prior stages are excluded.

\begin{table}[t]
    \centering
    \caption{\textbf{Incremental statistics of four datasets.} 
For each update period (week/day/block), we report the counts of users and items, the numbers of new users and new items, total interactions, and average actions per user/item. 
To simulate continual learning, datasets are partitioned into incremental stages: DIGINETICA by week, YOOCHOOSE by day, Gowalla from Feb. 2009 to Oct. 2010, and Amazon Sports from May 1996 to Oct. 2018. 
These datasets cover diverse temporal spans and sequence lengths, capturing varied user behaviors for robust evaluation.}
    \label{tab:dataset_stats}%
    \resizebox{\textwidth}{!}{%
    \rmfamily%
    \renewcommand{\arraystretch}{0.86}%
    \setlength{\tabcolsep}{4.7pt}%
    \begin{tabular}{@{}c|l|c|c|c|c|c|c|c|c|c}
    \toprule[1.2pt]
    \multirow{16}{*}{\makebox[15pt][c]{\rotatebox{90}{\textbf{DIGINETICA}}}}    & \cellcolor{gray!40}\textbf{Week} & \cellcolor{gray!40}\textbf{0} & \cellcolor{gray!40}\textbf{1} & \cellcolor{gray!40}\textbf{2} & \cellcolor{gray!40}\textbf{3} & \cellcolor{gray!40}\textbf{4} & \cellcolor{gray!40}\textbf{5} & \cellcolor{gray!40}\textbf{6} & \cellcolor{gray!40}\textbf{7} & \cellcolor{gray!40}\textbf{8} \\ 
    & \textbf{Users} & 15,076 & 7,896 & 6,453 & 6,870 & 6,379 & 11,996 & 11,043 & 12,113 & 14,301 \\ 
    & \textbf{New Users} & 15,076 & 7,801 & 6,393 & 6,818 & 6,316 & 11,943 & 10,954 & 12,000 & 14,298 \\ 
    & \textbf{Items} & 18,569 & 14,142 & 12,774 & 13,274 & 12,957 & 19,104 & 18,228 & 19,560 & 21,355 \\ 
    & \textbf{New Items} & 18,569 & 4,123 & 273 & 192 & 176 & 208 & 125 & 81 & 42 \\ 
    & \textbf{Interactions} & 70,739 & 37,586 & 31,089 & 32,687 & 30,419 & 57,913 & 52,225 & 57,100 & 69,042
 \\ 
    & \textbf{Avg Actions/User} & 4.69 & 4.76 & 4.82 & 4.76 & 4.77 & 4.83 & 4.73 & 4.71 & 4.83 \\ 
    & \textbf{Avg Actions/Item} & 3.81 & 2.66 & 2.43 & 2.46 & 2.35 & 3.03 & 2.87 & 2.92 & 3.23 \\  
    & \cellcolor{gray!40}\textbf{Week} & \cellcolor{gray!40}\textbf{9} & \cellcolor{gray!40}\textbf{10} & \cellcolor{gray!40}\textbf{11} & \cellcolor{gray!40}\textbf{12} & \cellcolor{gray!40}\textbf{13} & \cellcolor{gray!40}\textbf{14} & \cellcolor{gray!40}\textbf{15} & \cellcolor{gray!40}\textbf{16} & \cellcolor{gray!40}\textbf{Total} \\ 
    & \textbf{Users} & 16,885 & 16,613 & 9,999 & 12,639 & 14,091 & 15,166 & 11,835 & 15,982 & 204,789 \\ 
    & \textbf{New Users} & 16,881 & 16,613 & 9,996 & 12,634 & 14,090 & 15,163 & 11,835 & 15,978 & - \\ 
    & \textbf{Items} & 23,291 & 22,798 & 17,810 & 19,946 & 21,431 & 21,380 & 18,423 & 21,167 & 43,136 \\ 
    & \textbf{New Items} & 28 & 10,368 & 3,006 & 2,302 & 1,800 & 999 & 500 & 344 & - \\ 
    & \textbf{Interactions} & 82,834 & 82,935 & 50,037 & 63,133 & 70,050 & 71,670 & 56,959 & 77,065 & 993,483 \\ 
    & \textbf{Avg Actions/User} & 4.91 & 4.99 & 5.00 & 5.00 & 4.97 & 4.73 & 4.81 & 4.82 & 4.85 \\ 
    & \textbf{Avg Actions/Item} & 3.56 & 3.64 & 2.81 & 3.17 & 3.27 & 3.35 & 3.09 & 3.64 & 23.03 \\ 
    \midrule

    \multirow{16}{*}{\makebox[0pt][c]{\rotatebox{90}{\textbf{YOOCHOOSE}}}} 
    & \cellcolor{gray!40}\textbf{Day} & \cellcolor{gray!40}\textbf{0} & \cellcolor{gray!40}\textbf{1} & \cellcolor{gray!40}\textbf{2} & \cellcolor{gray!40}\textbf{3} & \cellcolor{gray!40}\textbf{4} & \cellcolor{gray!40}\textbf{5} & \cellcolor{gray!40}\textbf{6} & \cellcolor{gray!40}\textbf{7} & \cellcolor{gray!40}\textbf{8} \\ 
    & \textbf{Users} & 52,806 & 51,538 & 53,966 & 39,649 & 41,592 & 72,876 & 58,203 & 41,887 & 47,419 \\ 
    & \textbf{New Users} & 52,806 & 51,526 & 53,929 & 39,636 & 41,576 & 72,858 & 58,180 & 41,886 & 47,398 \\ 
    & \textbf{Items} & 12,885 & 12,294 & 12,939 & 12,503 & 12,968 & 14,069 & 12,662 & 12,866 & 13,346 \\ 
    & \textbf{New Items} & 12,885 & 2,707 & 417 & 389 & 359 & 358 & 263 & 278 & 254 \\ 
    & \textbf{Interactions} & 219,389 & 209,219 & 218,162 & 162,637 & 177,943 & 307,603 & 232,887 & 178,076 & 199,615 \\ 
    & \textbf{Avg Actions/User} & 4.15 & 4.06 & 4.04 & 4.10 & 4.28 & 4.22 & 4.00 & 4.25 & 4.21 \\ 
    & \textbf{Avg Actions/Item} & 17.03 & 17.02 & 16.86 & 13.01 & 13.72 & 21.86 & 18.39 & 13.84 & 14.96 \\ \arrayrulecolor{black} 
    & \cellcolor{gray!40}\textbf{Day} & \cellcolor{gray!40}\textbf{9} & \cellcolor{gray!40}\textbf{10} & \cellcolor{gray!40}\textbf{11} & \cellcolor{gray!40}\textbf{12} & \cellcolor{gray!40}\textbf{13} & \cellcolor{gray!40}\textbf{14} & \cellcolor{gray!40}\textbf{15} & \cellcolor{gray!40}\textbf{16} & \cellcolor{gray!40}\textbf{Total} \\ 
    & \textbf{Users} & 44,130 & 30,074 & 34,978 & 69,521 & 65,379 & 47,007 & 38,831 & 26,424 & 816,019 \\ 
    & \textbf{New Users} & 44,087 & 30,074 & 34,970 & 69,513 & 65,352 & 47,004 & 38,820 & 26,404 & - \\ 
    & \textbf{Items} & 13,217 & 11,716 & 12,557 & 14,403 & 14,206 & 13,276 & 12,471 & 11,398 & 25,958 \\ 
    & \textbf{New Items} & 239 & 2,135 & 1,473 & 1,506 & 1,067 & 728 & 527 & 373 & - \\ 
    & \textbf{Interactions} & 179,889 & 123,750 & 153,565 & 300,830 & 259,673 & 187,348 & 154,316 & 105,676 & 3,370,578 \\ 
    & \textbf{Avg Actions/User} & 4.08 & 4.11 & 4.39 & 4.33 & 3.97 & 3.99 & 3.97 & 4.00 & 4.13 \\ 
    & \textbf{Avg Actions/Item} & 13.61 & 10.56 & 12.23 & 20.89 & 18.28 & 14.11 & 12.37 & 9.27 & 129.85 \\ 
    \midrule

    \multirow{8}{*}{\makebox[0pt][c]{\rotatebox{90}{\textbf{Gowalla}}}} 
    & \cellcolor{gray!40}\textbf{Block} & \cellcolor{gray!40}\textbf{D0} & \cellcolor{gray!40}\textbf{D1} & \cellcolor{gray!40}\textbf{D2} & \cellcolor{gray!40}\textbf{D3} & \cellcolor{gray!40}\textbf{D4} & \cellcolor{gray!40}\textbf{Total} & \cellcolor{gray!40}- & \cellcolor{gray!40}- & \cellcolor{gray!40}- \\ 
    & \textbf{Users} & 29,055 & 2,742 & 2,614 & 2,656 & 2,779 & 33,027 & - & - & - \\ 
    & \textbf{New Users} & 29,055 & 817 & 950 & 1,021 & 1,184 & - & - & - & - \\ 
    & \textbf{Items} & 66,190 & 4,786 & 4,629 & 4,762 & 4,826 & 70,760 & - & - & - \\ 
    & \textbf{New Items} & 66,190 & 952 & 1,091 & 1,246 & 1,281 & - & - & - & - \\ 
    & \textbf{Interactions} & 1,712,368 & 61,336 & 57,238 & 57,485 & 58,876 & 1,947,303 & - & - & - \\ 
    & \textbf{Avg Actions/User} & 58.94 & 22.37 & 21.90 & 21.64 & 21.19 & 58.96 & - & - & - \\ 
    & \textbf{Avg Actions/Item} & 25.87 & 12.82 & 12.37 & 12.07 & 12.20 & 27.52 & - & - & - \\ 
    \midrule

    \multirow{8}{*}{\makebox[0pt][c]{\rotatebox{90}{\textbf{Amazon Sports}}}} 
    & \cellcolor{gray!40}\textbf{Block} & \cellcolor{gray!40}\textbf{D0} & \cellcolor{gray!40}\textbf{D1} & \cellcolor{gray!40}\textbf{D2} & \cellcolor{gray!40}\textbf{D3} & \cellcolor{gray!40}\textbf{D4} & \cellcolor{gray!40}\textbf{Total} & \cellcolor{gray!40}- & \cellcolor{gray!40}- & \cellcolor{gray!40}- \\ 
    & \textbf{Users} & 504,634 & 49,927 & 45,975 & 47,689 & 43,241 & 642,366 & - & - & - \\ 
    & \textbf{New Users} & 504,634 & 36,324 & 34,050 & 35,216 & 32,142 & - & - & - & - \\ 
    & \textbf{Items} & 188,375 & 37,711 & 34,844 & 33,175 & 30,175 & 205,843 & - & - & - \\ 
    & \textbf{New Items} & 188,375 & 5,794 & 4,496 & 3,854 & 3,324 & - & - & - & -  \\ 
    & \textbf{Interactions} & 2,619,512 & 215,041 & 194,837 & 201,194 & 176,414 & 3,406,998 & - & - & -  \\ 
    & \textbf{Avg Actions/User} & 5.19 & 4.31 & 4.24 & 4.22 & 4.08 & 5.30 & - & - & -  \\ 
    & \textbf{Avg Actions/Item} & 13.91 & 5.70 & 5.59 & 6.06 & 5.85 & 16.55 & - & - & -  \\ 

    \bottomrule[1.2pt]
    \end{tabular}}%
\end{table}

\subsubsection{Baseline}
\label{sec:compared methods}

To rigorously evaluate \ourmodel, we compare it against several representative baselines widely used in incremental sequential recommendation, including replay-based methods (e.g., ADER) and regularization-based methods (e.g., EWC), as well as standard approaches such as fine-tuning, retraining, and linear models. These baselines cover a broad spectrum of strategies for retaining historical knowledge while learning from new data, providing a comprehensive context for assessing the effectiveness of our proposed framework. We deliberately restrict our study to attention-based sequential recommenders to control architectural confounders and isolate the effect of conflict-aware incremental learning: attention-only backbones are a dominant and standardized choice in modern SR, and focusing on them ensures that performance differences primarily reflect the incremental strategy rather than heterogeneous sequence encoders. Consequently, several recent methods—such as PISA~\citep{Yoo2025pisa}, SAIL-PIW~\citep{wang2023sailpiw}, ReLoop2~\citep{zhu2023reloop2}, and GraphSAIL~\citep{zhang2020graphsail}—are not included because they either depart from our attention-only setting or integrate additional architectural components beyond our scope. A further practical constraint is that these works do not provide publicly available implementations or sufficient experimental details, making faithful reproduction under consistent conditions challenging, especially for conflict-aware mechanisms. Instead, we focus on baselines for which reliable reproduction is feasible within the attention-based setting, enabling a fair and systematic assessment of \ourmodel. This setup ensures that the reported improvements can be attributed to the Fisher-guided conflict detection and selective knowledge filtering framework, and provides a meaningful reference for future research in conflict-aware incremental sequential recommendation.

\begin{itemize}
    \item \textbf{Scratch}: Train the model using only the current stage’s data without leveraging any previous knowledge. 

    \item \textbf{Fine-tuning}: The model trained at the previous stage is directly fine-tuned on the current stage’s data without considering old knowledge. This baseline reflects a straightforward adaptation strategy.
    
    \item \textbf{Retraining}: At each stage, the model is retrained from scratch on all available data up to that stage. While often strong, this approach is computationally costly and generally impractical for real-world continual learning.
    
    \item \textbf{LinRec}~\citep{Liu2023linrec}: A lightweight linear incremental recommender that tracks evolving user–item preferences with low compute and memory overhead.
    
    \item \textbf{EWC}~\citep{kirkpatrick2017overcoming}: A classic regularization approach that estimates parameter importance from past tasks, and penalizes updates that move away from those old-knowledge–critical parameters, thereby mitigating catastrophic forgetting.
    
    \item \textbf{ADER}~\citep{mi2020ader}: A replay-based continual recommender that maintains a memory of selected historical interactions (exemplars) and jointly trains with new data to preserve prior knowledge during incremental updates.
\end{itemize}

\subsubsection{Evaluation Metrics}
\label{sec:evaluation}

We adopt three widely used ranking-based metrics for sequential recommendation evaluation: \textbf{Recall@K}, \textbf{MRR@K}, and \textbf{NDCG@K}, where \(K \in \{10, 20\}\).  

Recall@K: Recall measures the proportion of ground-truth items that appear within the top-\(K\) ranked list predicted by the model. In sequential recommendation, the ground-truth is the next item the user actually interacted with. Formally, for a test case with ground-truth item \(y\) and a predicted ranked list \(\hat{Y}_K\) of length \(K\):
\begin{equation}
    \text{Recall@K} = \mathbb{I}(y \in \hat{Y}_K),
\end{equation}
where \(\mathbb{I}(\cdot)\) is the indicator function. Recall@K evaluates the model’s ability to place the correct next item within the top-\(K\) recommendations.

MRR@K (Mean Reciprocal Rank): MRR measures the ranking quality of the ground-truth item by computing the reciprocal of its rank, averaged over all test cases. For a test case, if the ground-truth item \(y\) is ranked at position \(r\), its reciprocal rank is \(1/r\). Restricting the evaluation within the top-\(K\) predictions, we define:
\begin{equation}
    \text{MRR@K} = 
    \begin{cases}
    \frac{1}{r}, & \text{if } r \leq K, \\
    0, & \text{otherwise}.
    \end{cases}
\end{equation}
This metric emphasizes not only whether the correct item appears in the recommendation list but also how highly it is ranked.

NDCG@K (Normalized Discounted Cumulative Gain): NDCG accounts for both the relevance and the position of the ground-truth item, giving higher scores to items ranked higher in the top-\(K\) list. For a test case with ground-truth item at rank \(r \le K\):
\begin{equation}
    \text{DCG@K} = \frac{1}{\log_2(r+1)}, \quad
    \text{NDCG@K} = \frac{\text{DCG@K}}{\text{IDCG@K}},
\end{equation}
where \(\text{IDCG@K}\) is the ideal DCG (maximum possible DCG) for normalization. NDCG@K thus reflects both the correctness and the ranking position of the recommendation.

Together, Recall@K, MRR@K, and NDCG@K provide a comprehensive assessment: Recall@K measures the coverage of relevant items, MRR@K emphasizes the rank of the correct next item, and NDCG@K further accounts for the position-weighted relevance within the top-\(K\) recommendations.

\subsubsection{Implementation Details}
\label{sec:details}
Our method is implemented using the PyTorch framework\footnote{\href{https://pytorch.org/}{PyTorch: https://pytorch.org/}}.
During training, the batch size is set to 256. 
We adopt the Adam optimizer with an initial learning rate of 0.0005 and momentum parameters $\beta_1=0.9$ and $\beta_2=0.98$. 
Each model is trained for up to 100 epochs, and early stopping is applied based on the validation performance of Recall@20, with a patience of 5 epochs. 
The backbone network follows a Transformer-based sequential recommendation architecture. 
Transformer models have become the de facto standard in sequential recommendation due to their strong capability in modeling long-term dependencies and capturing complex user–item interaction patterns efficiently. 
Specifically, our implementation is based on the architecture of SASRec~\citep{kang2018sasrec}, with a hidden dimension of 150, 2 Transformer blocks, and 1 self-attention head. 
The maximum sequence length is set to 80 for Gowalla and 50 for the other datasets; longer sequences are truncated accordingly.
To prevent overfitting, dropout regularization is applied, where the dropout rate is tuned in the range $[0, 0.3]$ using Optuna\footnote{\href{https://optuna.org/}{Optuna: https://optuna.org/}}. 
Similarly, the hyperparameter $\alpha$, which controls the trade-off between stability and plasticity, is searched in the range $[0, 4]$ through Optuna. 
All baselines use the same public datasets with unified temporal splits, k-core filtering, evaluation protocols, and hyperparameter search ranges; baselines are reproduced end-to-end locally to ensure strict comparability under the same splits, protocols, and backbone.

\begin{table}[t]
\centering
\caption{Overall performance comparison across all evaluated datasets, covering Recall, MRR, and NDCG. '*' denotes the best result, '\underline{    }' indicates the second-best result, and Imp shows the relative improvement of the best-performing model over all baselines.}
\label{tab:model_results}
\resizebox{\textwidth}{!}{%
    \rmfamily%
    \setlength{\tabcolsep}{3.5pt}
    \begin{tabular}{c|l|ccccccc|c} 
    \toprule[1.2pt]
    \multirow{2}{*}{\textbf{Dataset}} & \multirow{2}{*}{\textbf{Metric}} & \multicolumn{7}{c|}{\textbf{Method}} & \multirow{2}{*}{\textbf{Imp.(\%)}} \\ 
    \cline{3-9}&  & Scratch & Finetune & Retrain & LinRec & Ewc & ADER & \textbf{\ourmodel} \\ \midrule
    \multirow{6}{*}{DIGINETICA} 
     & MRR@10  & 0.1146 & 0.1369 & 0.1510 & 0.1516 & 0.1323 & \underline{0.1571} & \textbf{0.1595}* & {\color{gray}+1.528$\uparrow$} \\
     & Recall@10  & 0.2395 & 0.3020 & 0.3349 & 0.3315 & 0.2902 & \underline{0.3502} & \textbf{0.3590}* & {\color{gray}+2.513$\uparrow$} \\
     & NDCG@10  & 0.1440 & 0.1756 & 0.1940 & 0.1937 & 0.1693 & \underline{0.2023} & \textbf{0.2062}* & {\color{gray}+1.928$\uparrow$} \\
     & MRR@20  & 0.1193 & 0.1433 & 0.1588 & 0.1593 & 0.1385 & \underline{0.1649} & \textbf{0.1679}* & {\color{gray}+1.787$\uparrow$} \\
     & Recall@20  & 0.3067 & 0.3941 & 0.4478 & 0.4424 & 0.3807 & \underline{0.4634} & \textbf{0.4800}* & {\color{gray}+3.582$\uparrow$} \\
     & NDCG@20  & 0.1610 & 0.1989 & 0.2225 & 0.2217 & 0.1922 & \underline{0.2309} & \textbf{0.2368}* & {\color{gray}+2.555$\uparrow$} \\ \midrule
    
    \multirow{6}{*}{YOOCHOOSE} 
     & MRR@10  & 0.3372 & 0.3467 & 0.3455 & 0.3467 & 0.3489 & \underline{0.3525} & \textbf{0.3552}* & {\color{gray}+0.766$\uparrow$} \\
     & Recall@10  & 0.5925 & 0.6126 & 0.6114 & 0.6112 & 0.6148 & \underline{0.6240} & \textbf{0.6321}* & {\color{gray}+1.298$\uparrow$} \\
     & NDCG@10  & 0.3983 & 0.4103 & 0.4090 & 0.4099 & 0.4125 & \underline{0.4175} & \textbf{0.4214}* & {\color{gray}+0.934$\uparrow$} \\
     & MRR@20  & 0.3423 & 0.3521 & 0.3511 & 0.3522 & 0.3543 & \underline{0.3580} & \textbf{0.3609}* & {\color{gray}+0.810$\uparrow$} \\
     & Recall@20  & 0.6649 & 0.6889 & 0.6904 & 0.6899 & 0.6914 & \underline{0.7014} & \textbf{0.7119}* & {\color{gray}+1.497$\uparrow$} \\
     & NDCG@20  & 0.4167 & 0.4297 & 0.4291 & 0.4299 & 0.4319 & \underline{0.4371} & \textbf{0.4417}* & {\color{gray}+1.052$\uparrow$} \\ \midrule
    
    \multirow{6}{*}{Gowalla} 
     & MRR@10  & 0.3758 & 0.4069 & \underline{0.4642} & 0.4557 & 0.4178 & 0.4534 & \textbf{0.4685}* & {\color{gray}+0.926$\uparrow$} \\
     & Recall@10  & 0.5431 & 0.5735 & \underline{0.6620} & 0.6611 & 0.5946 & 0.6548 & \textbf{0.6697}* & {\color{gray}+1.163$\uparrow$} \\
     & NDCG@10  & 0.4166 & 0.4475 & \underline{0.5122} & 0.5056 & 0.4608 & 0.5023 & \textbf{0.5174}* & {\color{gray}+1.015$\uparrow$} \\
     & MRR@20  & 0.3788 & 0.4100 & \underline{0.4682} & 0.4597 & 0.4212 & 0.4574 & \textbf{0.4722}* & {\color{gray}+0.854$\uparrow$} \\
     & Recall@20  & 0.5866 & 0.6179 & \underline{0.7206} & 0.7186 & 0.6433 & 0.7135 & \textbf{0.7235}* & {\color{gray}+0.402$\uparrow$} \\
     & NDCG@20  & 0.4276 & 0.4587 & \underline{0.5270} & 0.5202 & 0.4732 & 0.5171 & \textbf{0.5311}* & {\color{gray}+0.778$\uparrow$} \\ \midrule
    
    \multirow{6}{*}{\shortstack{Amazon\\Sports}} 
     & MRR@10  & 0.1603 & \underline{0.1679} & 0.1600 & 0.1599 & 0.1614 & 0.1661 & \textbf{0.1700}* & {\color{gray}+1.251$\uparrow$} \\
     & Recall@10  & 0.1907 & \underline{0.2000} & 0.1867 & 0.1867 & 0.1914 & 0.1995 & \textbf{0.2038}* & {\color{gray}+1.900$\uparrow$} \\
     & NDCG@10  & 0.1676 & \underline{0.1755} & 0.1663 & 0.1662 & 0.1685 & 0.1741 & \textbf{0.1780}* & {\color{gray}+1.425$\uparrow$} \\
     & MRR@20  & 0.1613 & \underline{0.1692} & 0.1610 & 0.1609 & 0.1625 & 0.1674 & \textbf{0.1713}* & {\color{gray}+1.241$\uparrow$} \\
     & Recall@20  & 0.2045 & \underline{0.2188} & 0.2015 & 0.2018 & 0.2075 & 0.2184 & \textbf{0.2232}* & {\color{gray}+2.011$\uparrow$} \\
     & NDCG@20  & 0.1711 & \underline{0.1802} & 0.1701 & 0.1701 & 0.1726 & 0.1788 & \textbf{0.1829}* & {\color{gray}+1.498$\uparrow$} \\ 
    \bottomrule[1.2pt]

\end{tabular}}
\end{table}

\begin{table}[t]
\centering
\caption{
Total training time to convergence (minutes), peak memory usage (MB), and overall performance comparison across datasets. 
Note that ``Time'' reports the wall-clock time required for the model to reach optimal performance, rather than per-epoch duration.
``Memory'' refers to the system RAM/storage overhead required for maintaining historical data or model states, distinct from GPU memory (VRAM).
The Model Comparison columns report the relative percentage change computed as 
$(\ourmodel - Baseline)/Baseline \times 100\%$, averaged across all four datasets. 
For Time and Memory, negative values (↓) indicate lower consumption (better efficiency), 
while positive values (↑) indicate higher consumption. 
For Recall@20 (Perf.), positive values (↑) indicate performance improvement, 
and negative values (↓) indicate performance decrease.
}
\label{tab:resource_results_comparison}

\resizebox{\textwidth}{!}{%
    \rmfamily%
    \renewcommand{\arraystretch}{1}
    \setlength{\tabcolsep}{3pt}%
    
    \newcolumntype{C}[1]{>{\centering\arraybackslash}m{#1}}%

    \begin{tabular}{l|cc|cc|cc|cc| C{1.1cm} C{1.1cm} C{1.1cm} }%
    \toprule[1.2pt]

    \multirow{2}{*}{\textbf{Method}} 
    & \multicolumn{2}{c|}{\textbf{DIGINETICA}} 
    & \multicolumn{2}{c|}{\textbf{YOOCHOOSE}} 
    & \multicolumn{2}{c|}{\textbf{Gowalla}} 
    & \multicolumn{2}{c|}{\textbf{Amazon}} 
    & \multicolumn{3}{c}{\textbf{Model Comparison(\%)}} \\ 

    \cmidrule{2-3} \cmidrule{4-5} \cmidrule{6-7} \cmidrule{8-9} \cmidrule{10-12}

    & Time & Mem & Time & Mem & Time & Mem & Time & \multicolumn{1}{c|}{Mem}
    & $\Delta$Time & $\Delta$Mem & $\Delta$Perf. \\ 
    \midrule

    Finetune & 34  & 1,475  & 116  & 1,552  & 12  & 1,509  & 32  & 1,463  
    & {\color{gray}+51.0$\uparrow$} & {\color{gray}+18.0$\uparrow$} & {\color{gray}+11.4$\uparrow$} \\

    Retrain  & 172 & 1,707  & 380  & 2,301  & 211 & 2,277  & 311 & 1,836 
    & {\color{gray}-72.7$\downarrow$} & {\color{gray}-12.8$\downarrow$} & {\color{gray}+3.8$\uparrow$} \\

    LinRec   & 105 & 1,732  & 220  & 2,250  & 110 & 2,304  & 243 & 1,911 
    & {\color{gray}-56.8$\downarrow$} & {\color{gray}-13.7$\downarrow$} & {\color{gray}+4.2$\uparrow$} \\

    EWC      & 32  & 1,527  & 60   & 1,613  & 14  & 1,567  & 57  & 1,652 
    & {\color{gray}+79.8$\uparrow$} & {\color{gray}+11.3$\uparrow$} & {\color{gray}+11.2$\uparrow$} \\

    ADER     & 180 & 91,803 & 178  & 56,749 & 68  & 82,047 & 216 & 52,736 
    & {\color{gray}-46.9$\downarrow$} & {\color{gray}-97.5$\downarrow$} & {\color{gray}+2.0$\uparrow$} \\

    \ourmodel & 63 & 1,597 & 138 & 1,670 & 17 & 1,937 & 75 & 1,874 
    & \color{gray}-- & \color{gray}-- & \color{gray}-- \\

    \bottomrule[1.2pt]
    \end{tabular}%
}
\end{table}

\subsection{Overall Performance and Resource Efficiency Comparison (RQ1)}

Table~\ref{tab:model_results} and ~\ref{tab:resource_results_comparison} summarize the overall recommendation performance, training time, and peak memory usage of all methods on four datasets. \ourmodel consistently achieves the best performance across all metrics while maintaining efficiency comparable to lightweight regularization-based methods. On average, it improves Recall@20 by 2.0\% while reducing memory usage by 97.5\% and training time by 46.9\% compared to the strongest baseline(ADER). These results demonstrate that \ourmodel effectively balances accuracy and efficiency, achieving robust long-term adaptation in dynamic recommendation scenarios.

Scratch, which trains only on the current stage without leveraging historical data, generally exhibits the poorest performance. Retrain typically surpasses Finetune by using all accumulated data, but at the cost of substantially higher training time and memory. In contrast, \ourmodel, despite using only new-stage data, outperforms Retrain by efficiently acquiring new knowledge while mitigating forgetting. Although our method introduces additional per-step computation (e.g., Fisher calculation and dual-model forward passes), the total training time remains comparable to Finetune. This efficiency gain is attributed to faster convergence: by selectively filtering conflicting gradients and focusing updates on compatible knowledge, \ourmodel reaches optimal performance in fewer epochs, effectively offsetting the increased per-epoch cost.

Compared with regularization-based methods such as EWC, \ourmodel offers more flexible and knowledge-aware adaptation. EWC relies on fixed parameter importance estimated solely from the old model, which may indiscriminately preserve outdated or conflicting knowledge. In contrast, \ourmodel estimates Fisher information on the new-stage data to screen conflicts and selectively mask incompatible parameters, thereby preserving only compatible historical signals. Coupled with an InfoNCE-based consistency loss that aligns the updated model with the filtered reference, SA-CAISR achieves stable retention and fast adaptation, maintaining accuracy under distribution shifts without replay.

Relative to replay-based methods such as ADER, \ourmodel attains higher recommendation accuracy without maintaining large historical buffers, thereby sharply reducing memory usage and training latency. By combining Fisher-guided selective preservation of compatible knowledge with InfoNCE-based consistency for stable adaptation, \ourmodel delivers state-of-the-art performance under continual updates while remaining buffer-free and privacy-friendly—making it well-suited for large-scale, dynamic recommendation settings.

\subsection{Robustness under Varying Stage Intervals and Preference Shifts (RQ2)}

Across all datasets, \ourmodel demonstrates consistently strong robustness under different stage intervals and preference shift magnitudes. On average, it improves Recall@20 by 0.4–3.6\%, MRR@20 by 0.8–1.8\%, and NDCG@20 by 0.8–2.6\% over the best-performing baselines across all datasets, highlighting its superior ability to effectively balance historical knowledge retention and adaptation to evolving user preferences.

For DIGINETICA and YOOCHOOSE, where stages are short and shifts are mild, historical interactions remain highly relevant and replay methods (e.g., ADER) are competitive. Even so, \ourmodel yields additional gains by using Fisher-guided masking to retain compatible signals and InfoNCE-based consistency to assimilate new information without disrupting past knowledge. For Amazon Sports, characterized by long stage intervals and substantial preference shifts, historical samples become less representative of current user interests. Consequently, replay-based ADER tends to retain outdated information, whereas Finetune adapts rapidly but suffers from severe forgetting. In contrast, \ourmodel selectively suppresses parameters identified as conflicting via Fisher-guided estimation, while preserving representations compatible with historical knowledge. This enables efficient adaptation to new-stage data without compromising previously learned information, leading to superior performance compared with ADER.
For Gowalla, which exhibits intermediate temporal gaps where historical and emerging interests coexist, ADER’s limited replay coverage constrains adaptation, while Retrain benefits from full data at the cost of high computation. \ourmodel surpasses both  by dynamically filtering conflicting updates and enforcing representational consistency.

Overall, these results confirm that \ourmodel achieves robust long-term adaptation across diverse temporal dynamics by selectively filtering outdated knowledge and reinforcing consistent representations, effectively maintaining a balance between stability and adaptability.
\subsection{Impact of Fisher-guided Screening and InfoNCE Consistency (RQ3)}
In this subsection, we perform ablation studies to evaluate the contributions of different components and design choices in \ourmodel.

\subsubsection{Impact of InfoNCE and Fisher Components}

\begin{table}[t]
\centering
\caption{
Ablation study of the InfoNCE and Fisher components across four datasets.
“\checkmark” denotes the inclusion of a component, while “–” indicates its absence.
When InfoNCE is disabled, the model defaults to KL-based regularization for aligning current and reference representations.
}
\label{tab:component_ablation}
\resizebox{\textwidth}{!}{
    \rmfamily%
    \setlength{\tabcolsep}{2.5pt}
    \begin{tabular}{cc|ccc|ccc|ccc|ccc}
    \toprule[1.2pt]
    \multirow{2}{*}{Fisher} & \multirow{2}{*}{InfoNCE} 
    & \multicolumn{3}{c|}{\textbf{DIGINETICA}} 
    & \multicolumn{3}{c|}{\textbf{YOOCHOOSE}} 
    & \multicolumn{3}{c|}{\textbf{Gowalla}} 
    & \multicolumn{3}{c}{\textbf{Amazon Sports}} \\ 
    & & M@20 & R@20 & N@20 
    & M@20 & R@20 & N@20 
    & M@20 & R@20 & N@20 
    & M@20 & R@20 & N@20 \\ 
    \midrule
    – & – & 0.1480 & 0.4323 & 0.2107 & 0.3526 & 0.7005 & 0.4328 & 0.4433 & 0.6932 & 0.5015 & 0.1597 & 0.2081 & 0.1706 \\
    – & \checkmark & 0.1585 & 0.4400 & 0.2208 & \underline{0.3584} & 0.7034 & 0.4379 & \underline{0.4656} & 0.7092 & \underline{0.5225} & 0.1637 & 0.2103 & 0.1741 \\
    \checkmark & – & \underline{0.1656} & \underline{0.4659} & \underline{0.2319} & 0.3582 & \underline{0.7054} & \underline{0.4381} & 0.4616 & \underline{0.7183} & 0.5216 & \underline{0.1703} & \underline{0.2204} & \underline{0.1814} \\
    \checkmark & \checkmark & \textbf{0.1679} & \textbf{0.4800} & \textbf{0.2368} & \textbf{0.3609} & \textbf{0.7119} & \textbf{0.4417} & \textbf{0.4722} & \textbf{0.7235} & \textbf{0.5311} & \textbf{0.1713} & \textbf{0.2232} & \textbf{0.1829} \\
    \bottomrule[1.2pt]
    \end{tabular}
}
\end{table}

As shown in Table~\ref{tab:component_ablation}, both the Fisher-guided and InfoNCE-based components are critical for achieving stable and adaptive continual recommendation.

Without our Fisher-guided mechanism, training on new-stage data tends to indiscriminately update parameters, allowing responses that conflict with the old model to overwrite compatible historical patterns—thereby inducing interference and forgetting. In contrast, \ourmodel computes Fisher on the new-stage data to quantify parameter-level conflict with the reference model and applies Fisher-weighted stochastic masking to suppress high-conflict parameters while retaining compatible ones. This conflict-aware screening yields a “filtered teacher” that emphasizes stable, non-conflicting signals.

When InfoNCE is removed, the model relies on KL-based regularization to align output distributions with the filtered reference. While KL enforces global distribution matching, it often fails to explicitly preserve fine-grained embedding structures or distinguish hard negatives—nuances that InfoNCE captures more effectively. By incorporating InfoNCE, SA-CAISR enforces contrastive consistency between the updated model’s representations and those from the conflict-filtered reference, thereby maintaining local semantic integrity and separation boundaries. This contrastive regularization effectively curbs representation drift and proves particularly beneficial in later stages with significant preference shifts, where simple distributional alignment (e.g., KL) is insufficient.

The combination of both components yields the best performance, demonstrating their complementary roles: Fisher-guided masking ensures stability by selectively mitigating destructive interference, while InfoNCE enhances plasticity by preserving semantically consistent embeddings. Together, they enable \ourmodel to robustly adapt to evolving user preferences while maintaining previously acquired knowledge.

\subsubsection{Impact of Different Fisher Update Strategies}

\begin{table}[t]
\caption{
Ablation study on different strategies for updating Fisher weights.
“\checkmark” indicates that the corresponding strategy is applied, while “–” indicates it is not.
Batch denotes dynamic updating of Fisher weights at each training batch; Smooth denotes applying a smoothing strategy to Fisher updates.
}
\label{tab:fisher_abation}
\resizebox{\textwidth}{!}{%
    \rmfamily%
    \setlength{\tabcolsep}{3.5pt}
    \begin{tabular}{cc|ccc|ccc|ccc|ccc}
    \toprule[1.2pt]
    \multirow{2}{*}{Batch} & \multirow{2}{*}{Smooth} 
    & \multicolumn{3}{c|}{\textbf{DIGINETICA}} 
    & \multicolumn{3}{c|}{\textbf{YOOCHOOSE}} 
    & \multicolumn{3}{c|}{\textbf{Gowalla}} 
    & \multicolumn{3}{c}{\textbf{Amazon Sports}} \\ 
    \cmidrule(lr){3-5} \cmidrule(lr){6-8} \cmidrule(lr){9-11} \cmidrule(l){12-14}
    & & M@20 & R@20 & N@20 
      & M@20 & R@20 & N@20 
      & M@20 & R@20 & N@20 
      & M@20 & R@20 & N@20 \\ 
    \midrule
    – & – & 0.1625 & 0.4589 & 0.2274 & 0.3583 & 0.7055 & 0.4376 & 0.4635 & 0.7034 & 0.5180 & 0.1694 & 0.2200 & 0.1771 \\
    \checkmark & – & \underline{0.1647} & \underline{0.4661} & \underline{0.2313} & \underline{0.3597} & \underline{0.7087} & \underline{0.4401} & \underline{0.4652} & \underline{0.7063} & \underline{0.5215} & \underline{0.1707} & \underline{0.2212} & \underline{0.1820} \\
    \checkmark & \checkmark & \textbf{0.1679} & \textbf{0.4800} & \textbf{0.2368} & \textbf{0.3609} & \textbf{0.7119} & \textbf{0.4417} & \textbf{0.4722} & \textbf{0.7235} & \textbf{0.5311} & \textbf{0.1713} & \textbf{0.2232} & \textbf{0.1829} \\
    \bottomrule[1.2pt]
    \end{tabular}
}
\end{table}

Table~\ref{tab:fisher_abation} compares strategies for applying Fisher-guided protection. Computing static Fisher weights once per stage offers baseline protection against destructive updates but fails to track evolving patterns within the stage. In contrast, updating Fisher weights dynamically at each batch allows the model to detect batch-specific conflicts arising from incoming data, yielding a more flexible stability–plasticity trade-off.

Applying a smoothing strategy during these updates further stabilizes the Fisher weights, preventing sudden fluctuations that could either overprotect or underprotect certain parameters.  
The combination of dynamic batch-wise updates with smoothing achieves the best performance across all datasets, confirming that this strategy effectively preserves critical historical knowledge while enabling efficient acquisition of new preferences.

\subsubsection{Ablation Study: Impact of Top-K Hard Negatives in InfoNCE}

\begin{figure}[!t]
    \centering
    \includegraphics[width=1\textwidth]{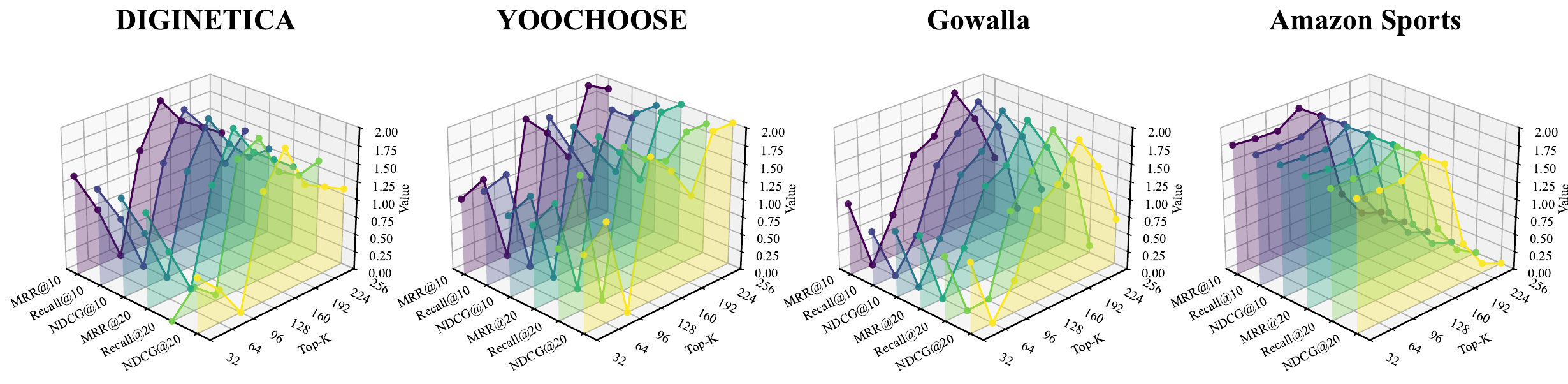}
    \caption{
    Effect of varying the Top-$K$ similarity threshold in the InfoNCE loss across four datasets.
    Each subfigure presents six normalized performance curves (MRR@10/20, Recall@10/20, NDCG@10/20) as $K$ increases.
    All metrics are individually normalized to emphasize their relative variation patterns rather than absolute values.
    }
    \Description{A diagram showing the ablation.}
    \label{fig:ablation_topk}
\end{figure}

To further examine the sensitivity of the InfoNCE loss to the selection of hard negatives, we vary the number of Top-$K$ hard negatives and report results in Figure~\ref{fig:ablation_topk}. Each subfigure corresponds to one dataset, and six curves represent the normalized trends of MRR@10/20, Recall@10/20, and NDCG@10/20 as $K$ increases. Since the absolute metric differences are relatively small compared to cross-metric scale variations, all metrics are independently normalized to better highlight their variation patterns with respect to $K$ rather than their absolute magnitudes.

From the overall trend, as $K$ increases from smaller values, model performance generally improves before gradually declining, although slight local fluctuations can be observed across some metrics. This indicates that incorporating a moderate number of hard negatives enhances the discriminative capacity of InfoNCE, allowing the model to better distinguish between compatible and conflicting knowledge representations. However, when $K$ becomes too large, the inclusion of less informative or easy negatives dilutes the learning signal, thereby interfering with stable representation alignment and weakening the balance between knowledge retention and new preference acquisition.

\subsubsection{Use-case Visualization Analysis}

\begin{figure}[!t]  
    \centering
    \includegraphics[width=1\textwidth]{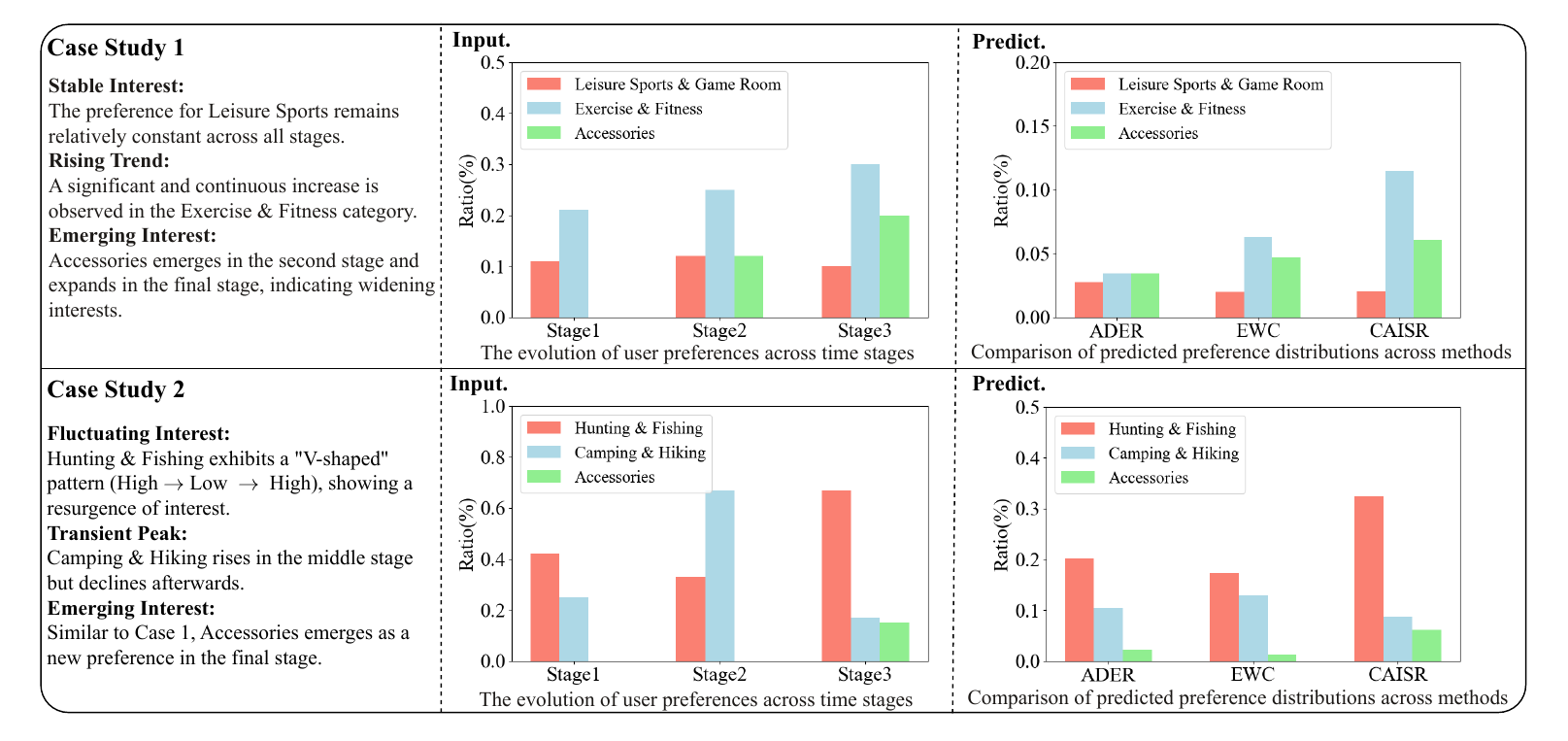}
    \caption{
    Illustrative example of model behavior under knowledge conflict. 
    When new data contradicts historical patterns, ADER and EWC tend to overfit outdated knowledge, leading to biased predictions. 
    In contrast, \ourmodel effectively identifies and filters conflicting information, enabling adaptive knowledge updating and more accurate recommendations in the new stage.
    }
    \Description{A diagram showing the case study.}
    \label{fig:case_study}
\end{figure}

To intuitively demonstrate the effectiveness of \ourmodel in handling complex preference drifts, we visualize the predicted probability distributions of different models against the ground-truth user evolution in Figure~\ref{fig:case_study}. In the first case, the user maintains a stable interest in \textit{Leisure Sports} while exhibiting a continuous rise in \textit{Exercise \& Fitness}. Notably, a new interest in \textit{Accessories} emerges in the intermediate stage and expands significantly in the final stage. For the stable category, all methods exhibit similar performance with minimal variance. However, for the rapidly growing \textit{Exercise \& Fitness} preference, \ourmodel achieves a significantly higher prediction ratio than ADER and EWC. This is because our method timely filters out conflicting historical knowledge that might hinder adaptation, allowing the model to swiftly align with the rising trend. Furthermore, regarding the \textit{Accessories} category, representative baselines tend to underestimate its importance due to the gradual accumulation of samples; in contrast, \ourmodel successfully identifies this as a non-conflicting new pattern and maximizes its preservation as it grows across stages.

The second case presents a more challenging scenario with fluctuating interests, where \textit{Hunting \& Fishing} follows a ``descent-then-ascent'' trajectory. Here, \ourmodel demonstrates superior sensitivity by accurately capturing the resurgence of interest in the final stage. In comparison, EWC fails to adapt sufficiently because the category was deemed unimportant in the previous low-interest stage, resulting in suppressed updates, while ADER's performance is limited by the averaging effect of replay. A similar advantage is observed in the \textit{Camping \& Hiking} category, which shows an inverse ``ascent-then-descent'' trend; \ourmodel correctly adjusts to the decline, whereas baselines struggle to discard the strong signal learned from the previous stage. Finally, for the \textit{Accessories} category which appears solely in the final stage, \ourmodel again effectively retains this fresh signal without it being diluted by historical data, validating its ability to balance stability and plasticity across diverse dynamic patterns.

In summary, these qualitative results provide empirical evidence that \ourmodel effectively resolves the critical stability-plasticity dilemma in incremental sequential recommendation. Unlike regularization-based methods that often suffer from historical inertia, or replay-based approaches that may lag behind rapid shifts due to the averaging effect of memory buffers, our framework excels in dynamic adaptation. By leveraging the Fisher-guided conflict screening mechanism, \ourmodel is able to precisely distinguish between compatible historical patterns and outdated noise. This capability allows the model to selectively preserve long-term preferences while aggressively adapting to emerging or fluctuating interests, ultimately ensuring that recommendations remain both accurate and timely in highly non-stationary environments.

\section{Conclusion}
In this work, we addressed the dual challenges of catastrophic forgetting and knowledge obsolescence in incremental sequential recommendation by proposing \ourmodel. Fundamentally distinct from existing replay-based and regularization-based paradigms, our approach neither relies on retrieving historical buffers nor imposes static constraints based solely on prior models. Instead, we theoretically reframe incremental adaptation as a dynamic conflict resolution process. By introducing a novel Fisher-weighted knowledge-screening mechanism, \ourmodel quantifies parameter-level conflicts between historical priors and emerging data distributions. This allows the model to selectively "forget" conflicting information while strictly preserving compatible patterns, offering a theoretically grounded solution to the stability-plasticity dilemma that traditional methods struggle to resolve.

From a practical perspective, \ourmodel demonstrates significant implications for real-world deployment. As a buffer-free framework, it naturally circumvents privacy concerns associated with data storage while achieving exceptional efficiency—reducing memory usage by 97.5\% and training latency by 46.9\% compared to state-of-the-art baselines. These characteristics make \ourmodel highly suitable for resource-constrained environments and large-scale systems requiring high-frequency updates. 

\section{Limitations and Future Work}

While \ourmodel establishes a robust buffer-free framework for incremental sequential recommendation, we identify specific limitations that contextualize our contributions and point toward future research directions.

\paragraph{Limitations}
First, regarding \textbf{architectural generalization}, we restricted our evaluation to attention-based backbones (SASRec) to rigorously isolate the efficacy of our conflict-aware mechanism. We did not extend \ourmodel to Graph Neural Networks (GNNs) or State Space Models (SSMs) in this study. This is due not only to the scarcity of reproducible implementations for recent baselines but also because adapting Fisher information estimation to graph topologies or recurrent states presents non-trivial theoretical challenges (e.g., structure-dependent conflicts) that warrant separate investigation.
Second, our method relies on an \textbf{independence assumption in conflict estimation} to ensure linear computational complexity suitable for high-frequency updates. Our Fisher-weighted screening assumes a diagonal Fisher Information Matrix, ignoring covariance between weights. In deep networks where knowledge is entangled across neurons, this diagonal approximation may lead to suboptimal masking decisions compared to full-matrix methods.
Third, concerning the \textbf{granularity of conflict adaptation}, while \ourmodel is "Stage-Adaptive" in capturing temporal distribution shifts, it remains "User-Agnostic" within each stage. The conflict estimation operates at the global parameter level using stage-wide statistics, potentially struggling to disentangle short-term noise from long-term interest drifts for users with highly heterogeneous behavior patterns.

\paragraph{Future Work}
To address these limitations, we propose three focused extensions.
(1) \textbf{Cross-Architecture Adaptation}: We plan to generalize the framework to heterogeneous architectures by developing specialized Fisher estimation protocols for GNNs (handling structural forgetting) and SSMs (handling long-state forgetting).
(2) \textbf{High-Order Conflict Modeling}: We aim to transcend the diagonal assumption by exploring efficient approximations of the full Fisher matrix, such as K-FAC. This would enable the detection of conflicts in the parameter correlation space, offering more precise protection of historical knowledge.
(3) \textbf{Instance-Aware Personalization}: We intend to introduce fine-grained, instance-aware conflict detection. By incorporating user-specific uncertainty or gradient coherence signals, future versions could dynamically adjust the masking strength per user, effectively separating temporary behavioral shocks from genuine preference evolution.


\bibliographystyle{ACM-Reference-Format}
\bibliography{reference}










\end{document}